\documentclass[aps, prb, twocolumn, showpacs, superscriptaddress]{revtex4-1}

\usepackage{graphicx}
\usepackage{amsmath}
\usepackage{amssymb}
\usepackage{physics}
\usepackage{hyperref}
\usepackage{comment}
\hypersetup{colorlinks=true,
	    final=true,
	    linkcolor=blue,
	    citecolor=blue,
	    filecolor=blue,
	    urlcolor=blue,
}

\usepackage{ulem}
\usepackage{xcolor}
\definecolor{orange}{RGB}{255,127,0}
\usepackage{makecell}

\newcommand\redsout{\bgroup\markoverwith{\textcolor{red}{\rule[0.5ex]{2pt}{0.4pt}}}\ULon}

\begin{document}
\title{Low Valence Nickelates: Launching the Nickel Age of Superconductivity}

\author{Antia S. Botana}
\affiliation{Department of Physics, Arizona State University, Tempe, AZ 85287, USA}
\email{Antia.Botana@asu.edu}

\author{Kwan-Woo Lee}
\affiliation{Division of Display and Semiconductor Physics, Korea University, Sejong 30019, Korea}

\author{Michael R. Norman}
\affiliation{Materials Science Division, Argonne National Laboratory, Lemont, IL 60439, USA}

\author{Victor Pardo}
\affiliation{Departamento de F\'{i}sica Aplicada, Facultade de F\'{i}sica, Universidade de Santiago de Compostela, E-15782 Spain}
\affiliation{Instituto de Materiais iMATUS, Universidade de Santiago de Compostela, E-15782 Spain}

\author{Warren E. Pickett}
\affiliation{Department of Physics and Astronomy, University of California Davis, Davis, CA 95616, USA}
\date{\today}

\begin{abstract}
The discovery  of superconductivity in thin films ($\sim$10 nm) of infinite-layer hole-doped NdNiO$_2$ has invigorated the field of high temperature superconductivity research, reviving the debate over contrasting views that nickelates that are isostructural with cuprates are either (1) sisters of the high temperature superconductors, or (2) that differences between nickel and copper at equal band filling should be the focus of attention. Each viewpoint has its merits, and each has its limitations, suggesting that such a simple picture must be superseded by a more holistic comparison of the two classes. Several recent studies have begun this generalization, raising a number of questions without suggesting any consensus. In this paper, we organize the findings of the electronic structures of $n$-layered NiO$_2$ materials ($n$= 1 to $\infty$) to outline (ir)regularities and to make comparisons with cuprates, with the hope that important directions of future research will emerge.
\end{abstract}

\maketitle

\section{Background}

\color{black}
After much synthesis and characterization of low-valence layered nickelates over three decades \cite{GREENBLATT1997174,hayward1999sodium,crespin1983reduced, crespin2005lanio2,poltavets2006la3ni2o6,kawai2010,ikeda2013}, superconductivity was finally observed \cite{Li2019_nature} in hole-doped $\cal{R}$NiO$_2$ (initially for rare earth $\cal{R}$=Nd, later for $\cal{R}$=La   \cite{ariando_La,Osada2021a_arxiv} and Pr \cite{Osada2020_nanolett}) with T$_c$ exhibiting a dome-like dependence \cite{Li2020_SC_dome,ariando2020prl} being maximal (10-15 K) near 20\% doping. %Their structure is identical to the `infinite-layer' cuprates ${\cal A}$CuO$_2$, with alkaline earth ${\cal A}$=Sr, Ca, though with a considerably lower  T$_c$.  
This series of discoveries in $\cal{R}$NiO$_2$ materials
marked the beginning of a new, nickel age of superconductivity \cite{trend,NRP} launching a plethora of experimental \cite{hwang_synthesis, goodge2020, deveraux, Fu2019, Li2020, BiXiaWang2020, QiangqiangGu2020} and theoretical \cite{Zhong_112_19,  Liu2020, arita, Thomale_PRB2020, Choi2020, Ryee2020, Gu2020, Leonov2020,lechermann,lechermann2, Hu2019, Sakakibara, jiang2020, werner2020, eff_ham,karp2020, prx, wang2020hund,olevano2020ab,vish2020prr,werner2020prx} work. 

%\color{blue} VP: I added a missing citation for hwang-synthesis, please check that it is the correct one. 

$\cal{R}$NiO$_{2}$ materials are the $n$=$\infty$ member of a larger series of layered nickelates with chemical formula ($\cal{R}$O$_2)^{-}$[$\cal{R}$(NiO$_2$)$_n$]$^{+}$ ($\cal{R}$=La, Pr, Nd; $n = 2, 3, \dots, \infty$) that possess $n$  cuprate-like NiO$_2$ planes in a square-planar coordination. Except for the $n$=$\infty$ case, groups of $n$-NiO$_2$ layers are separated by ${\cal R}_2$O$_2$ blocking layers that severely limit coupling between adjacent units.  These layered square-planar compounds are obtained via oxygen deintercalation from the corresponding parent perovskite ${\cal R}$NiO$_3$ ($n$=$\infty$)\cite{hayward1999sodium} and Ruddlesden-Popper  $\cal{R}$$_{n+1}$Ni$_n$O$_{3n+1}$ ($n$$\neq$$\infty$) phases \cite{GREENBLATT1997174}. As shown in Fig.~\ref{fig1}, the ($\cal{R}$O$_2)^{-}$[($\cal{R}$NiO$_2$)$_n$]$^{+}$ series can be mapped onto the cuprate phase diagram in terms of the nickel 3$d$-electron count, with nominal fillings running from $d^9$ ($n$=$\infty$) to $d^8$ (for $n$=$1$). 
That superconductivity arises in this series suggests that a new family of superconductors has been uncovered, currently with two members, $n$=$\infty$ and $n$=$5$,\cite{mundy} the only ones (so far) where an optimal Ni valence near $d^{8.8}$ has been attained. 

Some overviews on experimental and theoretical  findings in this family of materials have been recently published \cite{JunjieReview, review_arita, review_wen, review_cano}.
In this paper, we focus on the electronic structure of layered nickelates, confining ourselves to materials with the basic infinite-layer structure: $n$ square planar NiO$_2$ layers each separated by an ${\cal{R}}^{3+}$ ion without the apical oxygen ion(s) that are common in most cuprates and nickelates. 
%Except for the $n$=$\infty$ case, groups of $n$-NiO$_2$ layers are separated by R$_2$O$_2$ blocking layers that severely limit coupling of adjacent units.
%{\it [MRN - The rest of this paragraph needs a rewrite.] [VP: I would keep only the final part of the paragraph where the formal valence as a function of n is discussed, and I would also move here the discussion below about the I4/mmm space group characteristics]}
% The structural parameters are the tetragonal lattice parameters $a$ and $c$ and they have received much attention, especially noting that only strained samples, grown on mismatched substrates, have been found to superconduct. With the correspondence [Cu$^{2+}$,A$^{2+}$]$\rightarrow$ [Ni$^{1+}$,R$^{3+}$], the ionic sizes and transition metal (TM) site energies change, as well at the interionic hopping amplitudes. 
%The charged blocking layers (with the exception of the two extremes) lead to Ni formal valences that vary systematically with $n$ and that can be mapped into a prototypical cuprate phase diagram, as pictured in Fig.~\ref{fig1}. % This picture also indicates the regions of .....

%[We didn't mention magnetism much on zoom, but I presume that should be some part of this. The Stanford mafia has the recent magnetic excitations paper on $\infty$-layer nickelates that I must read more carefully. Basically long wavelength spinwaves without any LRO yet being identified. A measure of exchange coupling.]

\color{black}

\section{From $\infty$ to one}

\subsection{`Infinite-layer' $n=\infty$: ${\cal R}$NiO$_2$}

 %The $n=3$ members of the family (R$_4$Ni$_3$O$_8$) have been experimentally and theoretically shown to indeed be cuprate-like and promising  candidates for superconductivity if electron doping could be achieved. The higher-order Nd-based $n$=$5$ material (Nd$_6$Ni$_5$O$_{12}$), that falls directly into the cuprate dome area of filling without the need of doping, has also been recently found to be superconducting with a similar T$_c$$\sim$ 15 K. 

%AB commenting this out for now until we find a better spot for it 
%This is worth some discussion, especially given proposals in the literature that the superconductivity is an interfacial effect.\cite{ri2020prb,rossitza2021prr} For thin films, the $a$-axis is set by the substrate. Despite this, the quality of the films degrades once the thickness exceeds about 10 nm.  One reason is that as the thickness increases, there is a tendency to nucleate the $n=3$ phase instead given its more stable Ni valence.\cite{zeng2020phase}  As for bulk samples, since the precursor is cubic, there is no set orientation for the $c$-axis, meaning the bulk is far less ordered than the film.  There is also the possibility that the reduction agent (hydrogen) is trapped as well,\cite{onozuka2016formation}  although ab initio calculations suggest this is not stable for Sr-doped NdNiO$_2$\cite{si2020topotactic}.
 %And it should be remarked that several theoretical works have speculated that the superconductivity is an interface effect of the infinite-layer film with the STO substrate instead \cite{ri2020prb,rossitza2021prr}.

\begin{figure*}
	\centering
	\includegraphics[width=1.6\columnwidth]{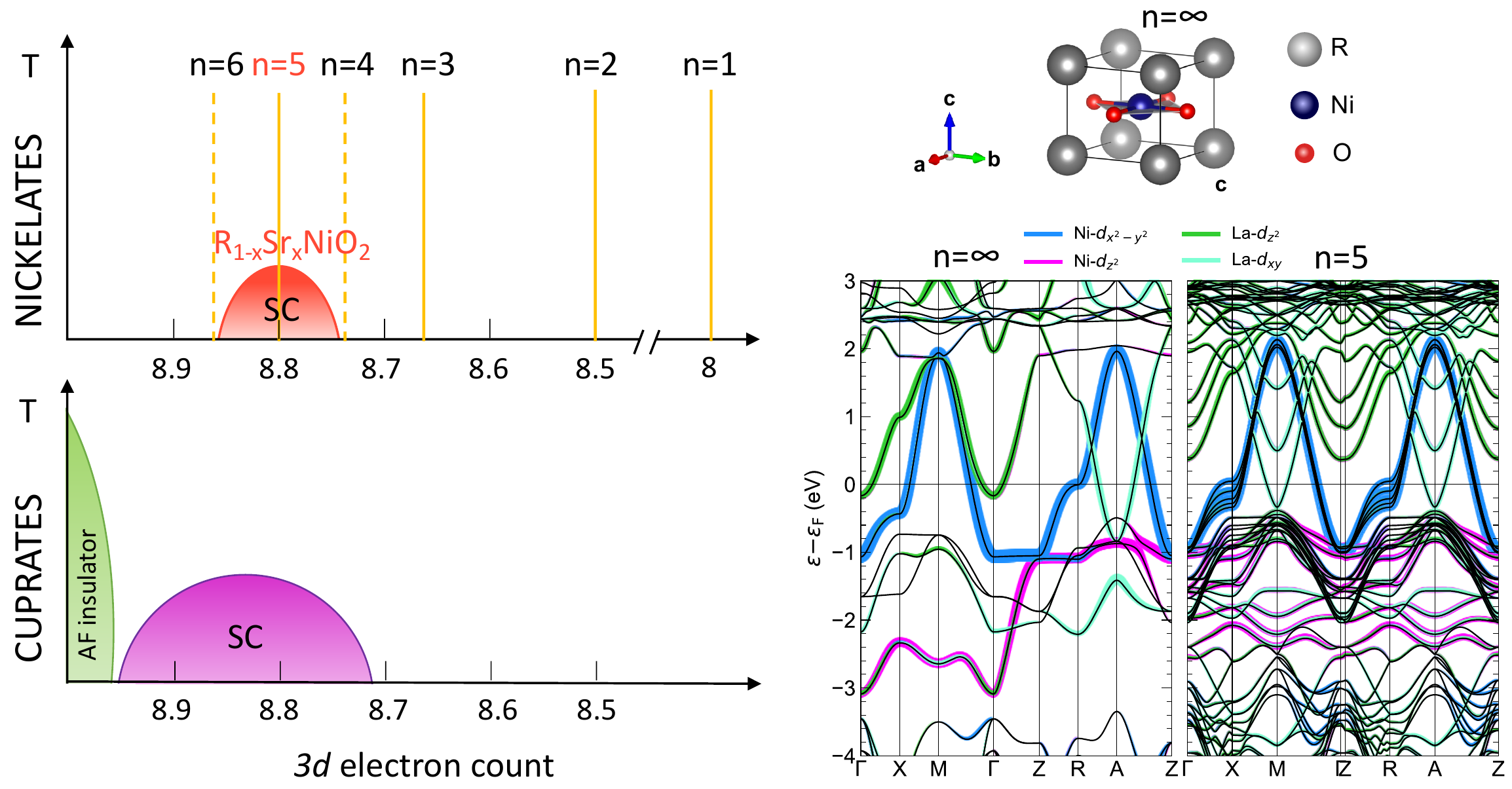}
	\caption{(Left) Schematic phase diagram as a function of nominal $d$ filling for layered nickelates (top) and cuprates (bottom), highlighting the regions where superconducting domes have been experimentally reported. The possible members of the $\cal{R}$$_{n+1}$Ni$_n$O$_{2n+2}$ series are marked with lines (dashed lines correspond to materials that have not been synthesized yet). The $n=5$ member falls close to the optimal doping value for both cuprates and the infinite-layer nickelate. (Top right) Basic structural unit of the infinite-layer $\cal{R}$NiO$_2$ with the square planar NiO$_2$ in the middle. (Bottom right) Band structures for the $n={\infty}$ and the $n=5$ materials, respectively. The two types of Ni $e_g$ bands are highlighted, as well as the two relevant $\cal{R}$-$5d$ bands.}
	\label{fig1}
\end{figure*}

In parent $\cal{R}$NiO$_2$ materials, 
%are similar to the cuprates in that their basic structural elements are NiO$_2$ square lattice planes, and 
Ni has the same formal 3$d^9$ electronic configuration as in cuprates. As mentioned above, superconductivity in $\cal{R}$NiO$_2$ materials emerges via hole doping, with T$_c$ exhibiting a dome-like dependence \cite{Li2020_SC_dome,ariando2020prl,osada2020prm} akin to cuprates, as shown in Fig.~\ref{fig1}. However, in parent
infinite-layer nickelates the resistivity shows a metallic T-dependence (but with a low temperature upturn)  \cite{ikeda2013,Li2019_nature} and there is no signature of long-range magnetic order, even though the presence of strong antiferromagnetic (AFM) correlations has recently been reported \cite{stanford2021}. This is in contrast to cuprates, where the parent phase is an AFM charge-transfer insulator. 

Noteworthy differences from cuprates were already reflected in early electronic structure calculations as well \cite{KWLee2004_prb,anisimov1999electronic}. For the parent material $\cal{R}$NiO$_2$, non-magnetic density functional theory (DFT) calculations show that besides the Ni-$d_{x^2-y^2}$ band, additional bands of $\cal{R}$-$5d$ character cross the Fermi level. The electronic structure of $\cal{R}$NiO$_2$ is three-dimensional-like, with a large $c$-axis dispersion of both (occupied) Ni and (nearly empty) $\cal{R}$-$5d_{z^2}$ bands due to the close spacing of successive NiO$_2$ planes along the $c$-axis. The $\cal{R}$-$5d_{z^2}$ dispersion leads to the appearance of electron pockets at the $\Gamma$ and A points of the Brillouin zone which display mainly  $\cal{R}$-$5d_{z^2}$ and $\cal{R}$-$5d_{xy}$ character, respectively, that self-dope the large hole-like Ni-$d_{x^2-y^2}$ Fermi surface. This self-doping effect (absent in the cuprates) introduces a substantial difference between nominal and actual filling of the Ni-$d$ bands, accounting for conduction and possibly also disrupting AFM order.  The presence of the 5$d$ electrons is consistent with experimental data, which reveal not only metallic behavior but also evidence for negative charge carriers as reflected in the negative Hall coefficient \cite{Li2019_nature,Li2020_SC_dome,ariando2020prl}. However, as the material becomes doped with Sr, the $\cal{R}$-$5d$ pockets become depopulated, the Hall coefficient changes sign\cite{Li2019_nature}, and the electronic structure becomes more single-band, cuprate-like \cite{prx, krishna2020effects}.

%WP \textit{(WP)  Analogies to cuuprates suggests an ACuO2 band panel to the left of RNiO2 in Fig. 1 would be useful. However, it complicates by needing more (at least a little more) discussion that U open the gap in the cuprates -- and that according to the Japanese group that calculated 6 spin configurations, a reasonable U also opens a (tiny) gap for (pi,pi,pi) order, the ground state. I believe this finally is due to moving down the Ni d bands relative to the R-5d bands -- the Ni-d bands are already Mott-gapped. See what I mean by requiring too much discussion?}

Besides the presence of $\cal{R}$-$5d$ electrons, infinite-layer nickelates have some other relevant differences from the cuprates, particularly their much larger charge-transfer energy between the metal $3d$ and oxygen $2p$ states. In cuprates, the charge-transfer energy  $\varepsilon_{3d}$-$\varepsilon_{2p}$ is as small as 1-2 eV \cite{Weber_2012}, indicative of a large $p$-$d$ hybridization, and enabling Zhang-Rice singlet formation. In $\cal{R}$NiO$_2$, the charge-transfer energy is much larger, $\sim$4.4 eV, as obtained from the on-site energies derived from a Wannier analysis \cite{prx}. This is consistent with the lack of a pre-peak in x-ray absorption data at the oxygen K-edge \cite{deveraux}. Because of this increase in charge transfer energy, the nickelate is more Mott-like, whereas the cuprate is more charge-transfer-like, in the scheme of Zaanen, Sawatzky and Allen \cite{jiang2020,karp2020}. Moreover, the doped holes tend to be on the Ni sites, as opposed to cuprates where they tend to reside on the oxygen sites. This in turn brings up the issue of the nature of the doped holes on the Ni sites.  That is, do
they behave as effective $d^8$ dopants, and if so, is $d^8$ high-spin or low-spin?  If the former, then these materials would fall in the category of Hund's metals \cite{wang2020hund,kang2020infinite,choi2021hund,hunds2021liu,kang2021hund}, and thus would deviate substantially from cuprates.  This important matter has yet to be resolved, though \textit{ab initio} calculations point towards a low-spin picture due to the large crystal-field splitting of the $e_g$ states in a square planar environment \cite{krishna2020effects}.
%However, holes in the form of Ni$^{1+}$ dopants are non-magnetic from DFT calculations, like in the cuprates.

Because of their lower degree of $p-d$ hybridization, the superexchange in $\cal{R}$NiO$_2$, as determined by resonant inelastic x-ray scattering experiments \cite{stanford2021}, is about half that of the cuprates.  Still, its value ($J$=64 meV) confirms the existence of significant AFM  correlations \cite{Leonov2020,stanford2021}. Long-range AFM order has however not been reported, with NMR data suggesting the ground state is paramagnetic \cite{zhao2021intrinsic},  and susceptibility data interpreted as spin-glass behavior \cite{lin2021universal}. Ne\'el type order
%with metallic band overlap
is consistently obtained in DFT studies \cite{KWLee2004_prb,Choi2020,Liu2020,Gu2020}, as in $d^9$ insulating cuprates.  %{\it [MRN - Really?  This would imply a sizable J2 interaction.]} [{\it [Kwan-woo's calculations gave AFM alignment clearly more stable than FM, and of course NM. Is that the question?  WEP}] [{\it -wep- The following sentences seem to me to need clarification, at least to me. The (small effects of Nd order need not be mentioned here.}] 
%This discrepancy between calculations and experiment could be either due to materials issues, to self-doped carriers (antiferromagnetism in cuprates is rapidly suppressed upon hole doping), or to difficulties in such measurements in RNiO$_2$ samples. 
The predicted AFM ground state in DFT+$U$ calculations\cite{zhang2021magnetic} 
%(and also in DFT+dynamical mean-field theory) 
is characterized by the involvement of both $d_{x^2-y^2}$ and $d_{z^2}$ Ni bands \cite{choi2020_prr}. This state is peculiar in that it displays a flat-band one-dimensional-like van Hove singularity of $d_{z^2}$ character pinned at the Fermi level. %This singularity makes the antiferromagnetic phase unstable to spin-density disproportionation, breathing lattice distortions, and charge-density disproportionation. 
These flat-band instabilities should inhibit
but not eliminate incipient AFM tendencies \cite{choi2020_prr}.
%{\it [MRN - I think we should be more inclusive here.  What is being referred to here appears to be a single DFT study.  We should comment on other DFT, DFT+U, and DMFT studies, as well as mention the NMR paper Antia referred to, and perhaps comment more on the recent RIXS paper in Science.]} {\it WEP - I also think these few sentences may be more detailed than necessary.}
%AB mention NMR exps above?

Discussing the origin of superconductivity in $\cal{R}$NiO$_2$, as in the cuprates, is a controversial topic.  But certainly the reduced T$_c$ of the nickelate compared to the cuprates is consistent with the reduced value of the superexchange, and the larger charge-transfer energy. $t$-$J$ model and RPA calculations building from tight-binding parameters derived from DFT calculations show that the dominant pairing instability is in the $d_{x^2-y^2}$ channel, as in cuprates \cite{wu2020robust}. Indeed, single-particle tunneling measurements on the superconducting infinite-layer nickelate have revealed a V-shape feature indicative of a $d$-wave gap \cite{tunneling}. On a broader level, several theoretical papers have speculated that the superconductivity is instead an interfacial effect of the infinite-layer film with the SrTiO$_3$ substrate\cite{ri2020prb,Bernardini_2020,rossitza2021prr,ortiz2021superlattice}, though recently superconductivity has been observed when other substrates are used \cite{Ren2021}. In this context, it should be noted that superconductivity has not been observed in bulk samples yet; since the precursor is cubic, there is no set orientation for the $c$-axis, meaning the bulk is far less ordered than the film \cite{Li2020,BiXiaWang2020}.

% VP: I've removed this comment since this is discussed at the beginning of the section. I moved the references there too. Finally, it should be remarked that several theoretical papers have speculated that the superconductivity is instead an interfacial effect of the infinite-layer film with the SrTiO$_3$ substrate \cite{ri2020prb,rossitza2021prr}.

%In addition, the increase in charge-transfer energy also suppresses the superexchange interaction, reducing the tendency for magnetism, even though according to calculations it does not cancel it altogether. 
%AB not sure if this can be controversial given the found 65 meV...
%This suppressed superexchange may also explain why the T$_c$ observed in these systems is fairly low as many theories for HTS in cuprates are precisely based on a small charge-transfer energy and large superexchange. 

\subsection{The superconducting $n=5$ material}

Recently, a second superconducting member has been found in the ($\cal{R}$O$_2)^{-}$[($\cal{R}$NiO$_2$)$_n$]$^{+}$ family: the   $n$=$5$ layered nickelate Nd$_6$Ni$_5$O$_{12}$, also synthesized in thin-film form \cite{mundy}. As schematically shown in Fig.~\ref{fig1}, this material has a nominal valence near that of the optimally-doped infinite-layer material (that is, Ni$^{1.2+}$: $d^{8.8}$ nominal filling) and so, in contrast to its infinite-layer counterpart, it is superconducting without the need for chemical doping. While $\cal{R}$NiO$_2$ displays NiO$_2$ layers separated by $\cal{R}$ ions, this quintuple-layer material (with five NiO$_{2}$ layers per formula unit) has an additional fluorite $\cal{R}$$_2$O$_2$ slab separating successive five-layer units. % (this structural difference is present in all of the RP-reduced phases for $n \neq \infty$, with $n$ the number of consecutive NiO$_2$ planes sandwiched by the fluorite slabs). \color{blue} [VP: I would move this phrase to the introduction where we discuss structural features, otherwise it may appear as it is something particular of the n=5 compound.] 
Further, each successive five-layer group is displaced by half a lattice constant along the $a$ and $b$ directions (i.e., the body centered translation of the $I4/mmm$ space group). These additional structural features effectively decouple the five-layer blocks, so the $c$-axis dispersion of this material is much weaker than its infinite-layer counterpart.
Despite these significant structural differences, T$_c$ is similar to that of the doped infinite-layer materials (with the onset of the superconducting transition taking place at $\sim$ 15 K), reducing the chances that yet to be synthesized low valence nickelates will have substantially higher transition temperatures.
%Also the R-Ni interactions are different, leading to smaller R-d bandwidths.

In terms of its electronic structure \cite{labollita2021electronic}, the $n$=$5$ material is intermediate between cuprate-like and $n$=$\infty$-like behavior. 
From DFT calculations, the charge-transfer energy of Nd$_6$Ni$_5$O$_{12}$ is $\sim$ 4.0 eV.
This reduced energy compared to the undoped infinite-layer material means that the Ni-$3d$ states are not as close in energy to the Nd-$5d$ states, consistent with the presence of a pre-peak in the oxygen K-edge (similar to what happens with Sr-doped NdNiO$_2$  \cite{krishna2020effects}).  As a consequence, the electron pockets arising from the Nd-$5d$ states are significantly smaller than those in the infinite-layer material (see Fig.~\ref{fig1}). This reduced pocket size along with the large hole-like contribution from the Ni-$3d$ states is consistent with experiment in that the Hall coefficient remains positive at all temperatures, with a semiconductor-like temperature dependence reminiscent of under- and optimally-doped layered cuprates. Aside from the appearance of these small Nd-derived pockets at the zone corners, the Fermi surface of Nd$_6$Ni$_5$O$_{12}$ is analogous to that of multilayer cuprates with one electron-like and four hole-like $d_{x^2-y^2}$ Fermi surface sheets. Importantly, the Fermi surface of the quintuple-layer nickelate is much more two-dimensional-like compared to the infinite-layer nickelate material, as the presence of the fluorite blocking slab reduces the $c$-axis dispersion, as mentioned above.
%The electronic structure hallmarks of the $n=5$ material indicate that it is more cuprate-like than its undoped infinite-layer counterpart, even though it also has some distinct features. 

\subsection{The $n$=$3$ material, the next superconducting member of the series?}

The materials discussed above can be put into the context of earlier studies of bulk reduced RP phases with $n$=$2$, 3 NiO$_2$ layers \cite{poltavets2006la3ni2o6, Poltavets2007_Ln4Ni3O8, zhang2017large,zhang2016stacked}, separated by fluorite $\cal{R}$$_2$O$_2$ blocking slabs that enforce quasi-2D electronic and magnetic behavior.

The $n$=$3$ member of the series, ${\cal R}_4$Ni$_3$O$_8$ (with Ni$^{1.33+}$: $d^{8.67}$ filling), has been studied extensively over the past decade (both single crystal and polycrystralline samples) \cite{Poltavets2007_Ln4Ni3O8}.
%AB add references
Since the charge-transfer energy decreases with decreasing $n$\cite{labollita2021electronic},
%and the ${\cal R}$=La member is insulating,
the $n$=$3$ class is more cuprate-like than its $n$=$\infty$ and $n$=$5$ counterparts. Both La and Pr materials are rather similar regarding their high-energy physics, with
a large orbital polarization of the Ni-$e_g$ states, so that the $d^{8}$ admixture is low spin \cite{pardo2012pressure,zhang2017large} (but see Ref.~[\onlinecite{karp-438}]). The primary difference is that La$_4$Ni$_3$O$_8$ exhibits long-range diagonal stripe order \cite{zhang2016stacked,zhang2019} (similar to that seen in 1/3 hole-doped La$_2$NiO$_4$), whereas its Pr counterpart appears to have short-range order instead \cite{lin2021strong}.  This results in the La material being insulating \cite{cheng2012pressure} in its low-temperature charge-ordered phase \cite{botana2016charge}, whereas Pr$_4$Ni$_3$O$_8$ remains metallic at all temperatures \cite{zhang2017large}, with an intriguing linear $T$ behavior in its resistivity for intermediate temperatures (similar to that of cuprates at a comparable hole doping).  Nd samples have also been studied \cite{retoux1998neutron}, but the degree of insulating/metallicity behavior seems to be sample dependent.  

The difference between La and Pr trilayer materials could be due to the reduced volume associated with Pr (one of the motivations for the authors of Ref.~[\onlinecite{Li2019_nature}] to study Sr-doped NdNiO$_2$ rather than Sr-doped LaNiO$_2$). The Ni spin state and metal versus insulator character have indeed been calculated to be sensitive to modest pressure \cite{pardo2012pressure}. Another factor is  possible mixed valency of Pr as observed in cuprates (though Pr-M edge data on Pr$_4$Ni$_3$O$_8$ did not indicate mixed valent behavior \cite{zhang2017large}).
Because of its decreased charge-transfer energy relative to $n=5$, the rare-earth derived pockets no longer occur \cite{pardo2010quantum}.
This lack of ${\cal R}$-$5d$ involvement is confirmed by the Hall coefficient that stays positive at all temperatures\cite{mundy}, (it remains to be understood why the thermopower in the case of La$_4$Ni$_3$O$_8$ is always negative \cite{cheng2012pressure}, also seen in the metallic phase). In addition, these trilayer nickelates show a reduced charge-transfer energy ($\sim$ 3.5 eV as obtained from a Wannier analysis \cite{labollita2021electronic}) that, along with the larger effective doping level, is consistent with the strong oxygen K edge pre-peak seen in x-ray absorption data \cite{zhang2017large}. Oxygen K edge RIXS data indicate a significant contribution of oxygen $2p$ states to the doped holes \cite{Shen2021}. As the effective hole doping level is 1/3, these materials are outside the range where superconductivity would be expected (see Fig.~\ref{fig1}).  Reaching the desired doping range for superconductivity might be possible via electron doping.  This could be achieved by replacing the rare earth with a 4+ ion (such as Ce or Th) \cite{botana2017electron}, intercalating with lithium, or gating the material with an ionic liquid.

If superconductivity were to occur, one might hope for a higher T$_c$ as has indeed been predicted via $t-J$ model calculations \cite{nica2020theoretical}. Recent RIXS measurements \cite{lin2021strong}, though, find a superexchange value for $n$=$3$ nearly the same as that reported for the infinite-layer material.  This suggests the possibility that T$_c$ in the whole nickelate family may be confined to relatively low temperatures compared to the cuprates.  The similar value of the superexchange for $n$=$\infty$ and $n$=$3$ is somewhat of a puzzle.  Though their $t_{pd}$ hoppings are very similar, the difference in the charge-transfer energy should have resulted in a larger superexchange for $n$=$3$.  The fact that it is not larger is one of the intriguing questions to be resolved in these low valence layered nickelates.

%Electron doping has been shown to be a plausible strategy to make these materials superconducting with (potentially) a higher T$_c$ than their infinite-layer counterparts, due to their reduced charge transfer energy and large superexchange. 

%AB add something on n=2

\subsection{The $n$=2 material}

The $n$=2 member of the series, La$_3$Ni$_2$O$_6$, has been synthesized and studied as well \cite{poltavets2006la3ni2o6,crocker2013nmr}.  In terms of nominal filling, it lies further away from optimal $d$-filling, being nominally Ni$^{1.5+}$: $d^{8.5}$. Experimentally, it is a semiconductor with no trace of a transition occurring at any temperature, although NMR data suggest that the AFM correlations are similar to those of the $n$=$3$ material. Electronic structure studies \cite{botana2016charge} have predicted its ground state to have a charge-ordered pattern with Ni$^{2+}$ cations in a low-spin state and the Ni$^{+}$: d$^9$ cations forming a S=1/2 checkerboard pattern. This charge-ordering between S=1/2 Ni$^{+}$: d$^9$ and non-magnetic Ni$^{2+}$: d$^8$ cations is somewhat similar to the situation in the $n$=$3$ material \cite{botana2016charge}. Calculations suggest that it is quite general in these layered nickelates that the Ni$^{2+}$ cations in this square-planar environment are non-magnetic. This has been shown by {\textit ab initio} calculations to be the case also with the Ni$^{2+}$ dopants in the ${\cal R}$NiO$_2$ materials \cite{RNiO2_nature_of_holes}.

\begin{comment}
{\it [MRN - We should talk about the next section.  It is a bit disjoint from the rest, with more details given than for the other materials.  Moreover, nothing is said about charge transfer physics, etc., again leading to the impression that this section is disjoint from the rest.  Besides, if we are going to talk about n=1, we should mention the 214 material, the n=2 reduced RP phase, as well as other materials like Sr2NiO2Cl2.  As I indicated above, it is not clear to me that the Ba2CuO3 material should be considered as near d8.  It appears to have d9-like regions buried in a d8-like matrix.  It is possible that the former is what is driving the superconductivity.]}
\end{comment}

\subsection{The $n$=1 case}
%MRN - a slight tweak to the beginning of this section
%{\color{blue} Instead of discussing t
 The long-known ${\cal R}_2$NiO$_4$ materials, with the $n$=1 formula as above, contain Ni ions with octahedral coordination.  We instead consider
Ba$_2$NiO$_2$(AgSe)$_2$ (BNOAS) \cite{bnoas19}, as it represents the extreme opposite of the $n$=$\infty$ member, not only in regards to its d$^8$ valence, but also because  its square planar coordination with long Ni-O bond is thought to promote `high-spin' (magnetic) behavior, that is, one hole in  $d_{x^2-y^2}$ and one hole in $d_{z^2}$. Unlike the other $n$ cases, the charge balanced formula is (BaAg$_2$Se$_2$)$^{0}$(BaNiO$_2$)$^{0}$; both blocking and active layers are formally neutral.
%{\it [MRN - This description of the magnetic susceptibility is confusing.]}
BNOAS is insulating, distinguished by a magnetic susceptibiliy that is constant, thus non-magnetic, above and below a peak at  T$^*$$\sim$130 K. This increase from and subsequent decrease to its high-T value  reflects some kind of magnetic reconstruction at T$^*$ that was initially discussed in terms of canting of high-spin moments. That interpretation does not account for the constant susceptibility above and below the peak. 
%With a small field applied, an enhanced constant susceptibility persists below the peak.
%This peculiar behavior, in the simplest interpretation, implies a paramagnetic material with singlets setting in below T$^*$=130 K.

Valence counting indicates Ni$^{2+}$: $d^8$, so a half-filled $e_g$ manifold. Conventional expectations are either (i) both $3d$ holes are in the $d_{x^2-y^2}$ orbital -- a magnetically dead singlet that cannot account for the behavior around T$^*$, or (ii) a Hund's rule $S$=1 triplet, which would show a Curie-Weiss susceptibility above the ordering temperature, but that is not seen in experiment. Correlated DFT calculations \cite{KWLee-etal} predict an unusual Ni $d^8$ singlet: a singly occupied $d_{z^2}$ orbital anti-aligned with a $d_{x^2-y^2}$ spin. This `off-diagonal singlet' consists of two fully spin-polarized $3d$ orbitals singlet-coupled, giving rise to a `non-magnetic' ion, however one having an internal orbital texture. Such tendencies were earlier noted \cite{KWLee2004_prb} in LaNiO$_2$, and related Ni spin states were observed to be sensitive to modest pressure in the $n$=2 and $n$=3 classes \cite{pardo2012pressure}. Attempts are underway\cite{tu2021iPEPS,singh2021bnoas} to understand this ``magnetic transition in a non-magnetic insulator''.

\color{black}
\section{Outlook}

While this new nickelate family seems to be emerging as its own class of superconductors, its connections to cuprates -- crystal and electronic structures, formal $d$ count in the superconducting region, AFM correlations -- retain a focus on similarities between the two classes.  Apart from the obvious structural analogy, the cuprate-motivated prediction of optimal $d^{8.8}$ filling has been realized in two nickelate materials, one achieved through chemical doping, the other layering dimensionality.
%One interesting point that was mentioned above is that the superconducting transition temperature of the $n=5$ material is similar to that of the infinite-layer materials, being significantly reduced compared to most cuprates.  This may be due to the reduced superexchange energy.  Although the superexchange value has not been determined for $n$=$5$, it has been for $n=3$, as we discuss in the next section.
In this context, the (so far) little studied $n$=$6$ and $n$=$4$ members of the series \cite{labollita2021electronic}
%if we look at the phase diagram in Fig.~\ref{fig1} and its amorphous boundaries (not shown),
may provide some prospect for superconductivity. Oxygen-reduced samples of these materials are so far lacking (though the $n$=$4$ member of the RP series has been epitaxially grown \cite{li2020epitaxial}), and even if they are synthesized, they might require additional chemical tuning to achieve superconductivity. They share a similar electronic structure to the $n=5$ material, but with slightly different nominal filling of the $3d$ bands \cite{labollita2021electronic}. Calculations show that as $n$ decreases from $n$=$\infty$ to $n$=$3$, the cuprate-like character increases, with the charge-transfer energy decreasing along with the self-doping effect from the rare earth $5d$ states. In contrast, the particular $n$=$1$ member discussed above seems distinct from other nickelates, and provides a different set of questions in the context of quantum materials \cite{tu2021iPEPS,singh2021bnoas}.

\section{Acknowledgments}
A.S.B. was supported by the U.S.~National Science Foundation, Grant DMR 2045826.
K.W.L.~was supported by the National Research Foundation of Korea, Grant No.~NRF2019R1A2C1009588. 
M.R.N.~was supported by the Materials Sciences and Engineering Division, Basic Energy Sciences, Office of Science, U.S.~Dept.~of Energy. 
V.P. acknowledges support from the MINECO of Spain through the project PGC2018–101334-BC21.  
W.E.P.~acknowledges support from U.S.~National Science Foundation, Grant DMR 1607139.

\end{document}